\documentclass{emulateapj}
\usepackage{amsmath,amssymb,graphicx,comment}
\usepackage{soul}
\usepackage[normalem]{ulem}

\usepackage{xcolor}

\shorttitle{The Black Hole Radio Background at Cosmic Dawn}
\shortauthors{Ewall-Wice et al.}

\begin{document}


\title{Modeling the Radio Background from the First Black Holes at Cosmic Dawn: Implications for the 21\,cm Absorption Amplitude}


\author{A. Ewall-Wice\altaffilmark{1,2}, T.-C. Chang\altaffilmark{1,3}, J. Lazio\altaffilmark{1}, O. Dor\'e\altaffilmark{1,3}, M. Seiffert\altaffilmark{1}, R. A. Monsalve\altaffilmark{4,5,6,7,8}}





\altaffiltext{1}{Jet Propulsion Laboratory, California Institute of Technology 4800 Oak Grove Dr, M/S 169-237, Pasadena CA 91109, USA}
\altaffiltext{2}{Dunlap Institute for Astronomy \& Astrophysics, 50 St. George St., Toronto, Ontario, M5S 3H4, Canada}
\altaffiltext{3}{California Institute of Technology, 1200 E California Blvd, Pasadena, CA 91125, USA}
\altaffiltext{4}{Department of Physics, McGill University, Montr\'eal, QC H3A 2T8, Canada}
\altaffiltext{5}{McGill Space Institute, 3550 Rue University, Montr\'eal, QC H3A 2A7, Canada}
\altaffiltext{6}{Center for Astrophysics and Space Astronomy, University of Colorado, Boulder, CO 80309, USA}
\altaffiltext{7}{School of Earth and Space Exploration, Arizona State University, Tempe, AZ 85287, USA}
\altaffiltext{8}{Facultad de Ingenier\'ia, Universidad Cat\'olica de la Sant\'isima Concepci\'on, Alonso de Ribera 2850, Concepci\'on, Chile}


\begin{abstract}
We estimate the 21\,cm Radio Background from accretion onto the first intermediate-mass Black Holes between $z\approx 30$ and $z\approx 16$. Combining potentially optimistic, but plausible, scenarios for black hole formation and growth with empirical correlations between luminosity and radio-emission observed in low-redshift active galactic nuclei, we find that a model of black holes forming in molecular cooling halos is able to produce a 21\,cm background that exceeds the Cosmic Microwave Background (CMB) at $z \approx 17$ though models involving larger halo masses are not entirely excluded. Such a background could explain the surprisingly large amplitude of the 21\,cm absorption feature recently reported by the EDGES collaboration.  Such black holes would also produce significant X-ray emission and contribute to the $0.5-2$\,keV soft X-ray background at the level of $\approx 10^{-13}-10^{-12}$\,erg\,sec$^{-1}$\,cm$^{-2}$\,deg$^{-2}$, consistent with existing constraints. In order to avoid heating the IGM over the EDGES trough, these black holes would need to be obscured by Hydrogen column depths of $ N_\text{H} \sim 5 \times 10^{23}\,\text{cm}^{-2}$. Such black holes would avoid violating contraints on the CMB optical depth from Planck if their UV photon escape fractions were below $f_{\text{esc}} \lesssim 0.1$, which would be a natural result of $N_\text{H} \sim 5 \times 10^{23}\,\text{cm}^{-2}$ imposed by an unheated IGM.
\end{abstract}

\section{Introduction}\label{sec:Intro}

The redshifted 21\,cm line of neutral Hydrogen (HI) offers a promising tool for mapping our universe's ``Cosmic-Dawn", when the first luminous sources formed (see \citealt{Barkana:2001,Furlanetto:2006,Morales:2010,Pritchard:2012,McQuinn:2016} for reviews). 

Several techniques are actively being pursued for detecting the 21\,cm signal. These include single dipole measurements of the sky-averaged ``global signal", which is being carried out by experiments such as EDGES \citep{Bowman:2010,Monsalve:2017}, SCI-HI \citep{Voytek:2014}, BIGHORNS \citep{Sokolowski:2015}, LEDA \citep{Bernardi:2016}, and SARAS \citep{Singh:2017}; and measurements of the power-spectrum of temperature fluctuations using antenna arrays. Interferometric experiments seeking to measure the power spectrum include the GMRT \citep{Paciga:2013}, the MWA \citep{Dillon:2014,Trott:2016,Jacobs:2016,Ewall-Wice:2016,Beardsley:2016}, PAPER \citep{Parsons:2014,Jacobs:2015,Ali:2015}, LOFAR \citep{Patil:2017}, and HERA \citep{DeBoer:2017,Kohn:2018}. An alternative technique for accessing the 21\,cm signal is to observe the absorption spectra of background radio sources (the ``21\,cm Forest") by the intergalactic medium (IGM) \citep{Furlanetto:2002,Carilli:2002,Furlanetto:2006Forest,Mack:2012,Ciardi:2013,Semelin:2016}. 

The EDGES collaboration has recently reported a detection of an absorption signature in the 21\,cm global signal centered at redshift $z\approx17$ \citep{Bowman:2018}. The most striking feature of this detection might be its nominal depth of 500\,mK, roughly twice as deep as what has been predicted by previous canonical models (see Fig 1. of \citet{Cohen:2017}). These models assume that the temperature of HI gas cannot cool below the adiabatic cooling limit for baryons decoupling from the Cosmic Microwave Background (CMB) at redshift $z\approx 150$ \citep{Peebles:1993} and that the CMB is the only significant 21\,cm background at early times.

The dependence of the amplitude of the absorption signal, on the radio background and gas temperature can be gleaned from the equation for the brightness temperature of 21\,cm radio emission/absorption from a distant HI cloud \citep{Madau:1997},
\begin{equation}\label{eq:Tb}
\delta T_b \propto \left( 1-\frac{T_{\text{CMB}}+T_{\text{rad}}}{T_s} \right),
\end{equation}
where $T_{\text{CMB}}$ and  $T_{\text{rad}}$ are the brightness temperatures of the CMB and any other 21\,cm background at the redshift of the HI cloud, and $T_s$ is the 21\,cm spin temperature. The amplitude of the absorption signal can be increased by reducing $T_s$. In canonical models for the 21\,cm signal, $T_s$ is primarily influenced by the Ly$-\alpha$ and X-ray backgrounds. The former couples $T_s$ to the kinetic temperature and the latter heats the HI. \citet{Barkana:2018}, \citet{Munoz:2018}, and \citet{Fialkov:2018} recently explored the possibility that the large amplitude of the absorption signal might be explained by the kinetic temperature being lowered through baryonic-dark matter interactions, first discussed by \citet{Tashiro:2014, Munoz:2015}. Alternatively, the amplitude of the absorption feature could be boosted, above previous expectations, by an additional radio background.

During the completion of this manuscript, \citet{Feng:2018} investigated the potential for 21\,cm experiments to constrain the  existence of a, so far, unexplained excess in the radio background measured by the ARCADE-2 experiment \citep{Fixsen:2011}. While the potential sources of this excess are varied, ranging from instrumental systematics to new astrophysics (see \citealt{Singal:2018} for an overview), \citet{Feng:2018} calculated that even a small fraction of the reported excess originating beyond $z\gtrsim 17$ can cause a very significant increase in the 21\,cm absorption feature amplitude. From Equation~\ref{eq:Tb}, we see that for fixed $T_s$, the presence of an additional back-light with temperature $T_{\text{rad}}$ will lead to a multiplicative increase in the absorption signal by a factor 
\begin{equation}\label{eq:Boost}
F_{\text{boost} } \approx 1+\frac{T_{\text{rad}}}{T_{\text{CMB}}},
\end{equation}
 when $T_{\text{rad}}+T_{\text{CMB}}\gg T_s$ where $T_s$ is the HI spin temperature. Thus, if a physical process can produce a radio background similar to or greater than the CMB, it can potentially explain the large absorption feature reported by \citet{Bowman:2018}. For instance, supernovae explosions of super-massive stars have been considered as a significant source of radio emission from $z\gtrsim 20$ \citep{Biermann:2014}.

It has been suggested that the presence of radio emission from active galactic nuclei (AGN) before or during the Cosmic Dawn might be detectable through its statistical imprint on the power-spectrum from 21\,cm forest absorption features \citep{Ewall-Wice:2014} or in its direct impact on the 21\,cm spin temperature \citep{Bolgar:2018}. Both \citet{Ewall-Wice:2014} and \citet{Bolgar:2018} used models that extrapolated empirical trends in nearby radio sources \citep{Haiman:2004,Wilman:2008} to predict the impact of these sources on the power-spectrum but neither of these works addressed the global signal. Given the dramatic amplitude of the EDGES feature, it is clearly worth exploring whether it might be produced by radio emission from growing black hole seeds.

In this paper, we explore the potential contribution to a high-redshift ($z \gtrsim 16$) radio background that might exist from accreting black hole seeds during the Cosmic Dawn. We derive our results using a simple analytic framework that only considers growth through continuous accretion. This simplified view should be valid at early times given the relatively massive birth halos (virial temperatures of $T_{\text{vir}}\gtrsim 10^3$\,K) that we consider \citep{Johnson:2013}. In \S\ref{sec:Modeling}, we present a simple motivational argument for the plausibility of black hole accretion producing the EDGES feature, before discussing our semi-analytic model for black hole-seed formation and growth. In \S\ref{sec:Results}, we present calculations of our model's contribution to the Cosmic X-ray Background (CXB), faint radio source counts, and the 21\,cm background experienced by HI during the Cosmic Dawn. We also explore the ionization and Lyman-Werner backgrounds that would be generated by these sources. In \S\ref{sec:Discussion} we discuss several of the issues that our model faces as an explanation for the reported EDGES signal, and conclude in \S\ref{sec:Conclusions}. Throughout this work, we assume the cosmological parameters from \citet{Planck:2016}.

\section{Modeling the Radio Emission from Early Black Hole Accretion}\label{sec:Modeling}

\subsection{Overall Motivation}
We begin with a simple calculation with which we demonstrate the plausibility of explaining the EDGES feature by radio emission from accreting black holes. 
The empirically motivated analysis that follows stems from three somewhat optimistic assumptions: (1) a non-negligible fraction of the known black hole mass already exists at high-redshift; (2) these black holes are undergoing Eddington to super-Eddington accretion; (3) these early black holes have radio-loudnesses similar to what is observed in active galactic nuclei (AGN) at $z\approx 0$. 

The emissivity of radio emission at redshift $z$ can be modeled as proportional to the mass density of black holes, $\rho_{bh}$,
\begin{equation}
\epsilon_\nu(z) \propto f_{\text{duty}}(z)f_{\text{edd}}(z)\rho_{\text{bh}}(z)
\end{equation}
where $f_{\text{edd}}(z)$ is the typical Eddington luminosity of black holes at redshift $z$, and $f_{\text{duty}}$ is the duty cycle. While black holes will be at significantly lower densities at high redshifts, their Eddington ratios and duty cycles may be significantly larger than the values found at $z \lesssim 1$. For example, \citet{Shen:2012} find that the typical Eddington ratios of broad line quasars increases from $\sim 10^{-2}$ at $z\approx 0$ to $f_{\text{edd}}\sim 0.3$ at $z \approx 4$. Meanwhile the duty cycle of black holes at $z \approx 0$ is $f_\text{duty} \sim 10^{-3}-10^{-2}$ \citep{Shankar:2009} while some models of black hole accretion in the early Universe find consitent $f_\text{duty} \sim 1$ for $\sim 10^8$ years \citep{Pacucci:2015}. For the sake of argument, we assume that a significant fraction (1\%) of the black hole mass has already been assembled between $z\approx 25$ and $z\approx 17$. We assume that each black hole emits in X-rays at some fraction, $f_X$, of the Eddington limit of $L_{\text{E}}=1.26 \times 10^{31}$\,W\,Hz$^{-1}$ $( \text{M}_{\text{bh}}/\text{M}_\odot)$. If $L_x\equiv f_X L_{\text{E}}=0.1 L_{\text{E}}$ is emitted in X-rays for each radio-loud black hole between $0.1-2.4$\,keV, we can assign a radio-luminosity according to the radio-quiet Fundamental plane in \citet{Wang:2006}, and boost the luminosity of radio-loud quasars ($f_L=10\%$ of the total population) by a factor of $10^R=10^{3}$ according to the typical radio loudness found in SDSS/FIRST AGN by \citet{Ivezic:2002}. If 1\% of the present-day black-hole mass ($\sim 10^4 h^2 \text{M}_\odot \text{Mpc}^{-3}$) was accreting at high redshift, we  obtain an emissivity,
\begin{align}
\epsilon(\nu,z) &\approx  1.2 \times 10^{22}\left(\frac{f_{L}}{0.1}\right)\left(\frac{f_{\text{duty}}}{1}\right)\left(\frac{10^{R}}{10^3}\right)\left(\frac{f_X}{0.1}\right)^{0.86} \nonumber \\ & \left(\frac{\rho_{\text{bh}}}{10^{4}h^2 \text{M}_\odot \text{Mpc}^{-3}}\right)\left(\frac{\nu}{1.4\text{GHz}}\right)^{-0.6} \text{W Hz}^{-1}h^3\text{Mpc}^{-3}.
\end{align}
An HI cloud at redshift $z$ would experience a specific intensity of
\begin{equation}\label{eq:Intensity}
J_\nu(z) = \frac{c(1+z)^3}{4\pi} \int_z^\infty \epsilon\left[\nu \frac{1+z'}{1+z},z^\prime\right]\frac{dz'}{(1+z')H(z')},
\end{equation}
where $H(z)$ is the Hubble parameter. Placing these sources between the redshifts $z=17$ and $z=25$, we compute the brightness temperature at redshift $z$ and $\nu=1420.41$\,GHz, $T_{\nu=1420.4\,\text{GHz}} \equiv T_{\text{rad}}(z)$, using 
\begin{equation}
T_\nu(z) =\frac{ c^2J_\nu(z)}{2 \nu^2 k_B}
\end{equation}
and find that $T_{\text{rad}}\approx 90$\,K, at $z=17$, nearly twice the CMB temperature at $z=17$.
Thus, with 1\% of the present day black-hole mass, we obtain a boost factor $F_{\text{boost}}\approx 3$, which is enough to explain the amplitude of the EDGES feature while leaving some room to spare for ionization and heating! We now proceed with a more physically motivated calculation of $\rho_{{\text{bh}}}$ during the Cosmic Dawn, where we consider a broad range of black hole seeding, growth, and emission scenarios. 

\subsection{Seeding Prescription}\label{ssec:Seeding}

How black holes quickly grew into the $10^8$-$10^9$\,$\text{M}_\odot$ Super Massive Black Holes (SMBH) observed only $\sim 1$\,billion years after the big bang remains a theoretical puzzle \citep{Fan:2003,Mortlock:2011,Wu:2015,Banados:2017}. Explanations for the progenitors of these super-massive black holes generally follow two paradigms.

In the first, black hole seeds with masses of $\sim 10-10^3\,\text{M}_\odot$ are formed as the remnants of the first generation of population III stars \citep{Madau:2001,Haiman:2001}. Such Pop-III stars are expected to form with masses of $\sim 10-10^3\,\text{M}_\odot$ \citep{Abel:2002,Bromm:2002,Yoshida:2008,Stacy:2010,Grief:2011,Hirano:2015} in molecular cooling halos with $T_{\text{vir}} \lesssim 10^4$\,K \citep{Haiman:1996,Abel:1997,Tegmark:1997}. In order for these black holes to reach SMBH masses by $z\sim 7$ requires the accretion to proceed very efficiently, i.e., at or above the Eddington limit \citep{Volonteri:2005,Volonteri:2006,Rhook:2006,Alexander:2014,Madau:2014,Volonteri:2015}.
 
 In the second, black holes form with initial masses of $\sim 10^4$--$10^5\,\text{M}_\odot$. These seeds can arise from ``Direct Collapse Black Holes" (DCBH)s \citep{Bromm:2003,Shang:2010,Johnson:2012}, self-gravitating pre-galactic disks \citep{Begelman:2006,Lodato:2006}, or runaway stellar mergers \citep{Devecchi:2009,Davies:2011,Stone:2017}. 
 
We capture the order-of-magnitude characteristics of seeding through these scenarios using the following semi-analytic model.  We model the evolution of a population of black hole seeds by constructing a grid of values of the \citet{Tinker:2008} halo-mass function between an initial redshift at which the seed halo density is nearly zero (we choose $z_\text{max}=80$) and $z_{\text{min}}=16$. At each time step, we assume that black hole seeds can form in some fraction, $f_{\text{halo}}$, of halos with masses (or virial temperatures) between some minimum~$\text{M}_{\text{min}}$ and maximum~$\text{M}_{\text{max}}$ values.
We assign each formation halo a black hole seed with mass $\text{M}_{\text{bh}}=f_{\text{seed}} \text{M}_{\text{halo}}$. Iterating forward in time, we estimate the number of new seeds added to the total black hole population at each time step to be $\Delta n_{\text{seed}}=\dot{n}_\text{seed} \Delta t$ where
\begin{eqnarray}
&\dot{n}_{\text{seed}}(\text{M}_{\text{bh,seed}})  =\nonumber\\
&\begin{cases} 
f_{\text{halo}} \dot{n}_{\text{halo}}(\frac{\text{M}_{\text{bh,seed}}}{f_{\text{seed}}}) & \quad \text{M}_\text{halo} \in [\text{M}_{\text{min}},\text{M}_{\text{max}}]\\
0 & \text{otherwise} 
\end{cases}
\end{eqnarray}
where $\dot{n}_{\text{halo}}$ is the number of halos that were below $\text{M}_{\text{min}}$ at $t-\Delta t$ and have accreted enough mass by time $t$ such that  $\text{M}_{\text{halo}}=\text{M}_\text{bh,seed}/f_{\text{seed}}\ge \text{M}_{\text{min}}$. We compute the accretion rate of halos at each time-step using the fitting formula of \citet{Fakhouri:2010},
\begin{equation}\label{eq:halo_accrete}
\dot{\text{M}}_{\text{halo}} = 46.1\,\text{M}_\odot \text{yr}^{-1}\left(\frac{\text{M}_{\text{halo}}}{10^{12}\text{M}_\odot}\right)^{1.1} (1+1.11z)H(z)/H(0).
\end{equation}
Table~\ref{tab:seeds} summarizes the parameters for three seeding models, which we now describe in more detail.

\begin{table}[h!]
\centering
\caption{Black Hold Seeding Model Parameters\label{tab:seeds}}
\begin{tabular}{ccccc}
Model & $f_{\text{seed}}$ & $f_{\text{halo}}$&$\text{M}_{\text{min}}$ &$\text{M}_{\text{max}}$\\ \hline
 Pop-III&$1.5\times10^{-4}$& $0.1$ &   $2.1 \times 10^3$ K & $10^4$ K \\
 DCBH&$10^{-2}$ & $10^{-4}$ & $10^7 \text{M}_\odot$ & $10^8 \text{M}_\odot$  \\
 Unstable Clusters & $10^{-3}$ & $5 \times 10^{-2}$ & $10^4$ K & $10^5$ K
\end{tabular}
\end{table} 

\subsubsection{ Pop-III Remnants}\label{sec:popiii}

We take Pop~III black hole seeds to form in halos having temperatures between $T_{\text{vir}}=2000$\,K and $T_{\text{vir}}=10^4$\,\hbox{K}. In principle, halos with masses down to several hundred K are capable cooling through H$_2$ and forming stars. However, the presence of baryonic-dark matter velocity offsets can increase this mass threshold by a factor of as much as $\approx 3$ \citep{Stacy:2011,Greif:2011b,Fialkov:2012}. We opt for a more conservative value of $\approx 2000$\,K also used by \citet{Tanaka:2016}. Our high-mass cutoff was chosen considering that the majority of atomic cooling halos are likely to have been formed from sub-halos that have already been metal enriched \citep{Johnson:2008}.

\citet{Tanaka:2016} consider models in which Pop-III seeds arise in $1$\% to 100\% of dark matter halos. We choose an intermediate value for the halo fraction, $f_{\text{halo}}\approx 0.1$.

What are our expectations for $f_\text{seed}$? The fraction of baryons that end up in stars within mini-halos is expected to be relatively low, and on average $f_\star \lesssim 10^{-3}$ \citep{Haiman:2006,Choudhury:2006,Visbal:2015}.  Meanwhile, stars with initial masses above $\gtrsim 240 \;\rm M_\odot$ are expected to directly collapse into high-mass black-holes \citep{Yoon:2012}. The fraction of stellar-mass that becomes locked into such objects depends primarily on the Pop-III initial mass function. Many studies find that the first proto-stellar disks fragment into small clumps \citep{Stacy:2010,Clark:2011,Greif:2011,Greif:2012} but it is still unclear whether these small fragments typically migrated to the center of the disk, forming a   $100-1000$\,M$_\odot$ star, or remained apart, forming much less massive objects. \citet{Hirano:2015} find that at $z \gtrsim 18$, before the EDGES feature, between $70$ and $100$\,\% of stars form above the direct collapse threshold (their Table~2) and while their 2-D simulations could not properly account for disk fragmentation, \citet{Hosokawa:2016} argue that fragmentation might actually increase final stellar masses by suppressing UV feedback through episodic accretion. We  adopt the more conservative  fraction of stellar matter that ends up in direct collapse black holes of $10$\%. Assuming $f_\text{halo}=0.1$, and $f_\star=10^{-3}$ across all halos with and without Pop-III formation  yields $f_\text{seed} \Omega_m / \Omega_b=10^{-3}$. Multiplying by the baryon-matter density ratio gives us $f_\text{seed} = 1.5 \times 10^{-4}$. 

The  assumption of $10$\,\% high mass black-hole formation might be relaxed by reducing the minimal virial temperature of halos. Lowering $T_\text{vir}^\text{min}$ to $500$\,K yields a potential seed host density at $z=25$ that is $\approx 31$ times greater than if the minimum mass is $2000$\,K so if we adopted this lower virial temperature and assumed only 0.3\% of Pop-III stars ended their lives by collapsing into high mass black-holes we would obtain similar black hole densities.

\subsubsection{ Direct Collapse Black Holes}\label{sec:dcbh}

We seed halos with masses between $10^7$ and~$10^{8}$~$\text{M}_\odot$ and assign a mass fraction of $f_{\text{seed}}=10^{-2}$, approximately reproducing the initial mass-function of \citet{Ferrara:2014} with seeds ranging from~$10^5\,\text{M}_\odot$ to~$10^6\,\text{M}_\odot$. The abundance of formation sites for DCBHs is currently uncertain, typically thought to require an unpolluted halo subject to a Lyman-Werner background. \citet{Dijkstra:2014} predict an occurrence rate of $\sim 10^{-7}\text{--}10^{-5} h^3 \text{Mpc}^{-3}$ at $z=10$ while \citet{Agarwal:2012} predicts significantly higher values of $\sim 10^{-3} h^3 \text{Mpc}^{-3}$. We use the fraction from \citet{Tanaka:2016} and set $f_{\text{halo}} \approx 10^{-4}$, which yields a formation site density of $\sim 10^{-3} h^3 \text{Mpc}^{-3}$ at $z=10$, similar to \citet{Agarwal:2012}'s scenario. It is possible that baryon-dark-matter velocity offsets might amplify the abundance of pristine atomic cooling halos and direct collapse holes \citep{Hirano:2017}, and that such objects may also be able to form in metal polluted halos \citep{Dunn:2018}. Hence the reader should consider our specific DCBH model as simply illustrative.

\subsubsection{ Unstable Clusters}\label{sec:clusters}

We include this model to represent the scenario presented by \citet{Devecchi:2009} in which mildly polluted atomic cooling halos, subject to a UV background, form dense clusters of stars in their cores. Runaway collisions in these clusters leads to the formation of $\sim 10^3\, \text{M}_\odot$ black hole seeds. \citet{Devecchi:2009} find that these seeds form in $f_{\text{halo}} \approx 5 \times 10^{-2}$ of halos above $T_{\text{vir}}= 10^4$K. To obtain a typical black hole mass of $1000\,\text{M}_\odot$, we set the mass fraction to be $10^{-4}$ of the host-halo mass. 

\subsection{Growth through Accretion}\label{ssec:Accretion}
After forming, we allow each seed to grow through accretion, radiating at 
$L = \frac{\eta}{1-\eta} \dot{\text{M}}_{\text{bh}}c^2$ at some fraction of the Eddington rate, $L = f_{\text{edd}} L_E(\text{M}_{\text{bh}})$, where $L_E = 1.26 \times 10^{31} \text{W} (\text{M}_{\text{bh}}/\text{M}_\odot) $, and $\eta$ 
is the radiative efficiency of accretion (the fraction of the infalling rest-energy that is radiated away). If the black hole accretes some $f_{\text{duty}}$ fraction of the time, the mass varies as \citep{Johnson:2013}
\begin{equation}\label{eq:Accrete}
\text{M}_{\text{bh}}(t) = \text{M}_{\text{bh,seed}} \exp \left(f_{\text{duty}} f_{\text{edd}} \frac{(1-\eta)}{\eta} \frac{t - t_{\text{seed}}}{\tau_E} \right),
\end{equation}
where $\tau_E\approx 0.45$~Gyr is the Eddington time-scale and $t_{\text{seed}}$ is time at which the seed formed. In order to keep our analysis simple, we assume that each seed grows continuously from accretion, which is a reasonable assumption for our model with seed halos of  $T_{\text{vir}}>10^3$K \citep{Johnson:2013}. We consider the range of radiative efficiencies discussed in \citet{Milosavljevic:2009}, of~0.025 to~0.4. Fixing the level of radio loudness to what is observed locally, we find that only efficiencies $\eta\lesssim0.1$ 
are capable of producing sufficient levels of radio emission.  We restrict our analysis to $0.03\le \eta\le 0.05$ 
in the following discussion (we will briefly revisit large $\eta$ 
 solutions in \S~\ref{sec:Discussion}). We also consider a range of $f_{\text{edd}}$ between $10^{-2}$ and $1$ and fix $f_{\text{duty}}=0.5$ which is an optimistic but reasonable assumption at high redshifts \citep{Shankar:2010}. 

\subsection{Radio Emission}\label{ssec:Emission}
We assume that each black hole has an intrinsic X-ray luminosity between 2 and 10\,keV given by $L_\text{2-10} = k_\text{bol}^{-1} f_\text{edd} L_\text{E}$ between 2  and 10\,keV, where we use a bolometric correction factor of $k_\text{bol}^{-1} = 0.07$, consistent with the distribution observed by \citet{Lusso:2010}. We then extrapolate to $L_X \equiv L_{0.1-2.4}$ by assuming that the X-ray luminosity follows a power law with index of $0.9$.

The nature of radio emission from accreting black-holes is still poorly understood. Thus, we will emply empirical trends.  The radio luminosity (associated with jet emission) of an accreting black hole, relative to its optical luminosity (associated with the accretion disk), is typically described by a {\it radio loudness} parameter, $R$, which either refers to the logarithm of the ratio between 5\,GHz and B~band (4250~\AA) 
luminosities \citep{Kellerman:1989} or 1.4\,GHz and i-band (8000 \AA)
ratios \citep{Ivezic:2002}. We adopt the latter definition. There is a lively debate as to whether the $R$ distribution is bi-modal, consisting of two physically distinct {\it radio quiet} and {\it radio loud} populations, where the latter sources are typically $\sim 1000$ times more luminous than radio quiet sources \citep{Ivezic:2002,Cirasuolo:2003, Ivezic:2004,Singal:2011}. There is also disagreement as to whether the radio-loudness distribution evolves over redshift and in which direction \citep{Jiang:2007, Donoso:2009,Singal:2011,Banados:2015}. Correlation between radio-emission and X-ray emission is also well documented, primarily in the ``fundamental-plane'' of black holes \citep{Merloni:2003} which, in most works, does not distinguish between radio-quiet and radio-loud sources. \citet{Wang:2006} find that radio-loud sources tend to lie far above the trends fitted with radio-quiet sources. 

 The task of this work is not to better understand our expectations for radio-loudness at high redshift but rather to investigate what plausible levels of radio loudness can produce the EDGES feature.  To this end, we adopt the bi-modal radio-loudness model of \citet{Ivezic:2002} and divide our black hole population into 10\% radio loud, and 90\% radio quiet, assuming that the radio-loud distribution does not evolve with redshift. Given $L_X$, we assign each black hole a radio luminosity at $1.4$ GHz using the radio-quiet fundamental plane \citep{Merloni:2003,Falcke:2004}  fitted in \citet{Wang:2006} for radio-quiet AGN: 
\begin{align}\label{eq:FP}
 L_{1.4}(\text{M}_{\text{bh}})&=\frac{8.3 \times 10^{-6}}{1.4\times10^9\text{Hz}} \left(\frac{L_X}{L_{\text{E}}(\text{M}_{\text{bh}})}\right)^{0.86}  L_{\text{E}}(\text{M}_{\text{bh}})\nonumber \\ &\times \begin{cases} 1 & \text{ radio quiet} \\ \langle 10^{0.83R} \rangle & \text{ radio loud }\end{cases},
\end{align}
where the multiplicative ratio for radio emission from radio loud and radio quiet AGN predicted by the fundamental plane scales as $\sim 10^{0.83R}$.  We calculate the average $\langle 10^{0.83R}\rangle$ radio-luminosity for our radio-loud population by modeling $e^R$ as log-normal distributed with a mean of $\langle R \rangle=2.8 \times \ln(10)$ and a standard deviation of $\sigma = 1.1 \times \ln(10)$, which describes the distribution of radio-loudness for SDSS/FIRST AGN at redshift $\approx 1$ (see Fig 19. of \citet{Ivezic:2002}). This radio loudness distribution yields a boost factor of $\langle 10^{0.83R} \rangle \approx 1.9 \times 10^3$ for radio loud AGN.

To match the spectral index of the observed radio background, we assign our radio sources a spectral index of $-0.6$. This value is within the range of what one would expect for synchrotron emitted by shock injected electrons, 
such as what is observed in the lobes and hot-spots of low-redshift radio sources. However, diffuse lobe emission may be difficult to produce at high redshifts due to enhanced Compton cooling from CMB photons \citep{Ghisellini:2014,Ghisellini:2015,Saxena:2017}. \citet{Sharma:2018} notes that the short cooling times at high redshift would lead to spectral aging and an index closer to $\approx -1.1$. Our computed radio background amplitude at $z \approx 18$ is not very sensitive to the choice of spectral index, although we note that the higher value suggested by \citet{Sharma:2018} would help our scenario obey radio source count constraints by making the sources observed at $z=0$, and higher rest-frame frequency appear fainter.

\section{Results}\label{sec:Results}
We now present our results. We start by presenting our model's contribution to the 21\,cm radio background (\S~\ref{ssec:Boost}). We then discuss the predictions our model makes for faint radio source counts (\S~\ref{ssec:RadioCounts}), the soft X-ray background (\S~\ref{ssec:Xrays}), the density of black holes between the redshifts $20$ and $16$ (\S~\ref{ssec:Density}), and our model's implications for reionization (\S~\ref{ssec:Ionization}) and Lyman-Werner feedback (\S~\ref{ssec:LWBackground}). 

\subsection{The Impact on Radio Background and 21\,cm Signal during the Cosmic Dawn}\label{ssec:Boost}
We compute the boost factor for the 21\,cm absorption trough (assuming no heating from  X-rays) given in Equation~\ref{eq:Boost}. This quantity provides us with a plausible upper limit on the amount by which our black holes can boost the 21\,cm signal. 

\begin{figure*}
\includegraphics[width=\textwidth]{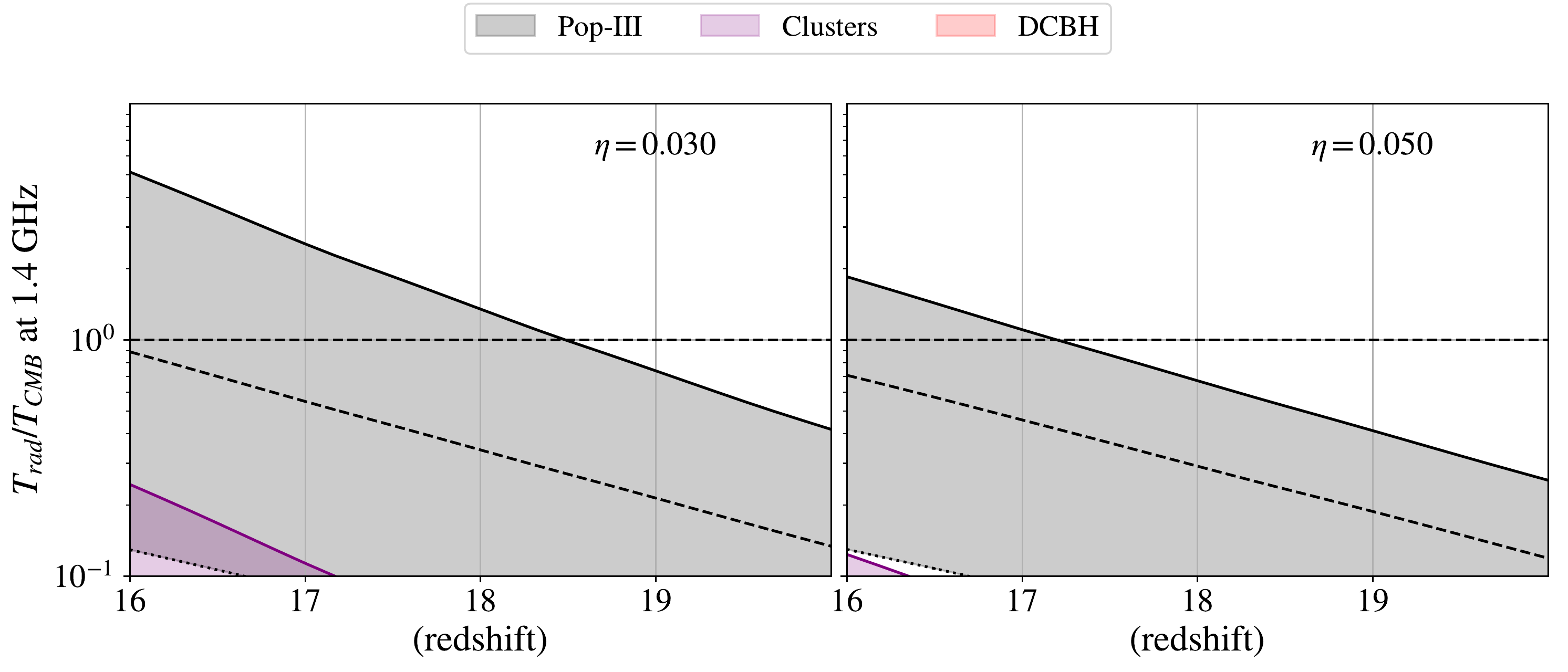}
\caption{
The ratio between $T_{\text{rad}}$  and $T_{\text{CMB}}$, as a function of redshift for $\eta=0.03$ (left) and $\eta = 0.05$ (right).
The filled shaded regions denote the range of temperatures predicted between $f_{\text{edd}}=10^{-1}$ and $1$ for our Pop-III (grey), DCBH (red), and Cluster collapse (purple) scenarios (see table~\ref{tab:seeds}). In the absence of heating, models that pass above the horizontal dashed line at $z \lesssim 17$ are producing a sufficient radio background to account for the absorption feature detected by the EDGES experiment.}\label{fig:Boost}
\end{figure*}

We plot $T_{\text{rad}}/T_{\text{CMB}}$, at 21\,cm, in Fig.~\ref{fig:Boost} 
for $\eta = 0.03$ and $\eta=0.05$ with filled regions denoting Eddington ratios between $f_{\text{edd}}=10^{-2}$ and $1$. It is apparent that amongst our black hole models, only our Pop-III scenario produces a radiative background close to the CMB and capable of producing the EDGES absorption feature. If a radiative background from accreting black holes is responsible for the observed 21\,cm absorption amplitude of $\approx 500$\,mK, then it is unlikely that these objects were constrained to form in halos above the atomic cooling threshold, at least under the assumptions of our DCBH and cluster models.

We conclude, from Fig.~\ref{fig:Boost}, that we can produce enough radio background to explain the EDGES feature while still satisfying the CXB constraints with Pop-III black holes that formed in 10\% of eligible halos before and during the Cosmic Dawn.
These black holes need to have accreted with a duty cycle of $f_{\text{duty}}\approx 0.5$ and radiative efficiency of $\eta\lesssim 0.05$
, while radiating with an Eddington efficiency between $f_{\text{edd}}\approx 0.1$ and $1$. These numbers can be relaxed if we allow for higher radio emission efficiencies or a larger 
radio-loud fraction. 

From Fig.~\ref{fig:Boost}, it is clear that DCBH and cluster collapse black holes cannot produce enough radio emission to explain the EDGES absorption feature unless they formed with significantly higher efficiencies than is theoretically expected. We find that for $f_L=0.1$, cluster collapse black holes approach necessary levels of radio emission if $f_{\text{halo}}\approx 0.5$ while DCBH's require $f_{\text{halo}}\approx 10^{-2}$. 

It is worth briefly discussing the contribution of our model to the radio background observed at $z=0$. Our most emissive model, $\eta=0.03, f_{\text{edd}}=1$ 
, produces a brightness temperature, at $1.4$\,GHz, of $\approx 8 T_{\text{CMB}}$ at $z=16$, roughly $371$\,K. At $z=0$, this would be observed as $21.84$\,K emission at $83$\,MHz. For our spectral index of $-0.6$ in flux density, this gives $T_{\text{rad}}(\nu, z=0)\approx 33\left(\nu/1\text{GHz}\right)^{-2.6}\,\text{mK}$, over an order of magnitude below the claimed detection of an $\approx 1$\,K excess observed by ARCADE-2 \citep{Fixsen:2011}. This result is consistent with \citet{Feng:2018}'s observation that background temperatures significantly below those measured by ARCADE-2 can explain a large increase in the absorption feature. 


\subsection{Radio Source Counts}\label{ssec:RadioCounts}
\begin{figure*}
 \centering
 \includegraphics[width=\textwidth]{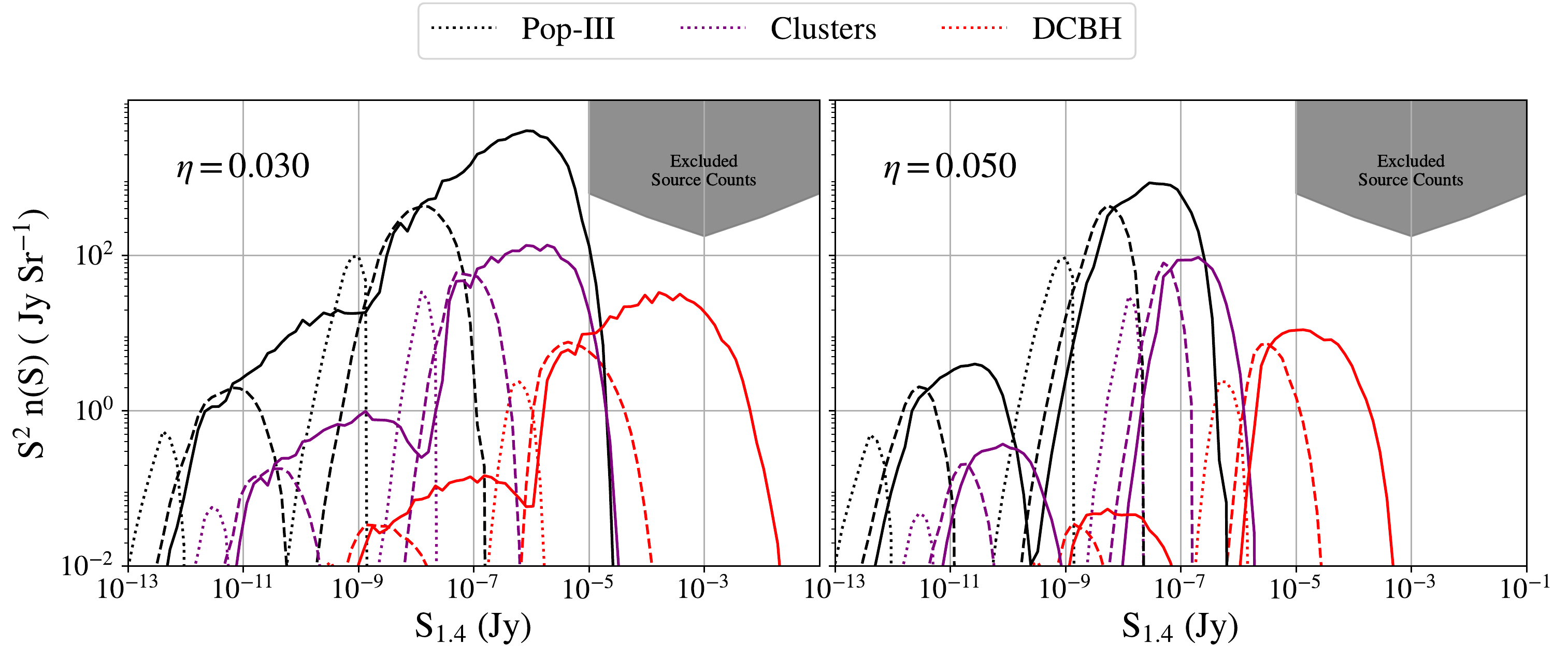}
 \caption{The number of sources per steradian per flux interval on the sky from $z\ge16$, times flux-squared, for our various models of black hole seeding and growth. Solid lines indicate $f_{\text{edd}}=1$ while dashed and dotted  
 lines assume $f_\text{edd} = 10^{-1}$ and $f_{\text{edd}}=10^{-2}$ respectively. Left Column: $\eta = 0.03$. Right Column: $\eta = 0.05$. The grey shaded region indicates flux counts excluded by \citet{Condon:2012}. All of our scenarios predict flux counts below existing limits.
 Reducing the radiative efficiency to $\approx 0.03$ leads to our $f_\text{edd} \approx 1$ scenario to violate existing point source constraints though a steeper spectral index might still allow for it.}
 \label{fig:Radio}
 \end{figure*}

 In Fig.~\ref{fig:Radio} we show the number of sources per flux interval and solid angle on the sky observed at $z=0$ from our radio sources at $z \ge 16$ at $1.4$ GHz. All of our models show two peaks in flux corresponding to the radio-loud and radio-quiet sub-populations. Low radiative efficiencies and high Eddington rates lead to larger populations of more massive black holes, forming high-flux wings. Our Pop-III models yield fluxes below the detection threshold of any existing surveys, $\lesssim 10^{-5}$ Jy, while our cluster and DCBH scenarios lead to $\sim$mJy sources that only contribute to the detected source counts at the $1-10$ percent level.  All scenarios considered yield populations of sources below the limits on source counts imposed by surveys (e.g. \citet{Condon:2012,Vernstrom:2014}).
 
\subsection{Soft X-ray Background}\label{ssec:Xrays}
\begin{figure}\centering
\includegraphics[width=.5\textwidth]{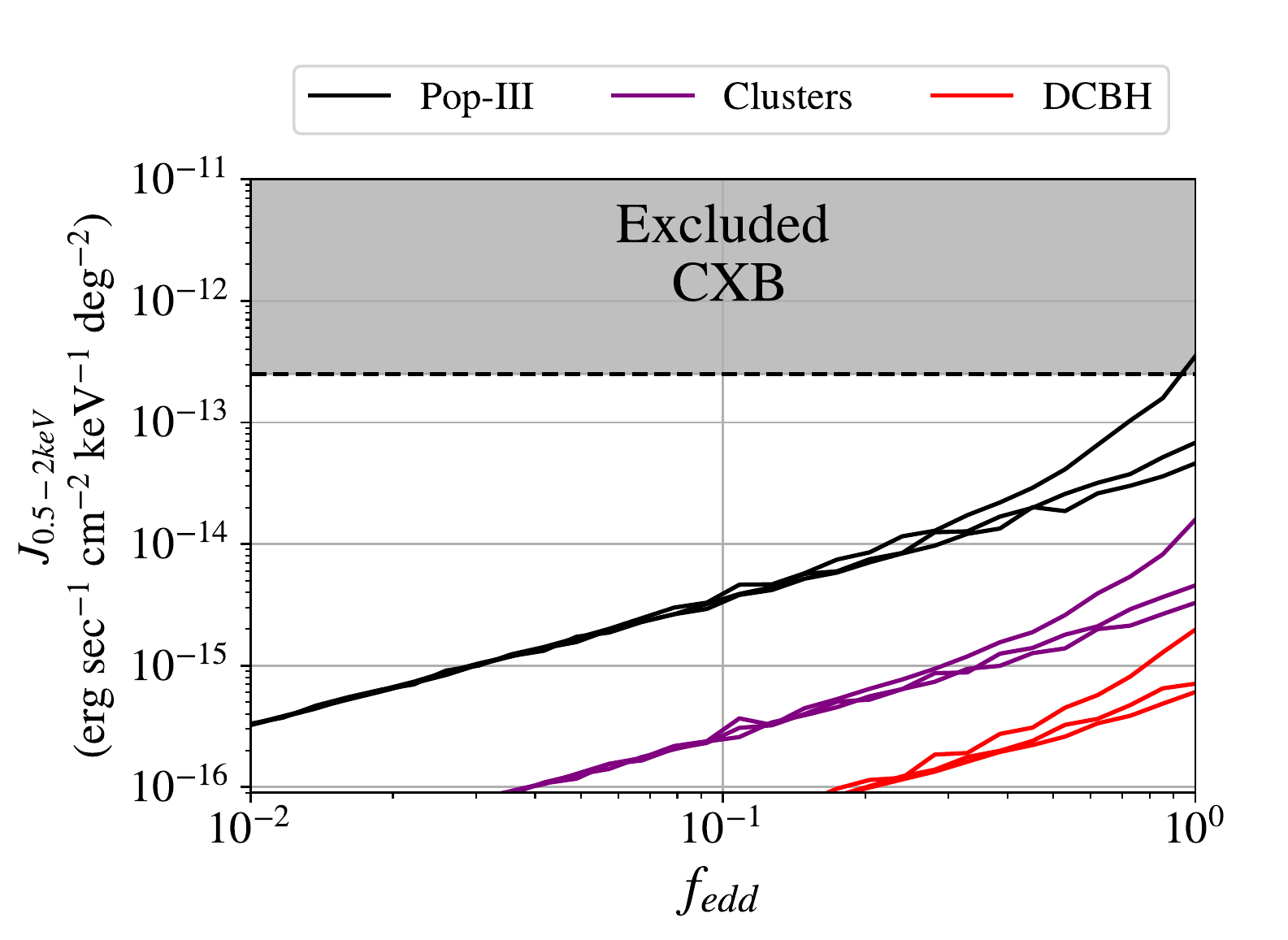}
\caption{The integrated X-ray background between 0.5 and 2 keV predicted from our black hole models assuming obscuration below $0.5$\,keV and a spectral index of $0.9$ for an accreting Black Hole population above $z=16$. Each set of lines for each seeding model corresponds to radiative efficiencies of $0.4$, $0.05$, and $0.03$ with background amplitudes increasing with decreasing radiative efficiency. All of our models fall below the X-ray background constraint from \citet{Fialkov:2017} (black dashed line).}\label{fig:Xrays}
\end{figure}

Our model also predicts the contribution from our accreting black holes to the Cosmic X-ray Background \citep{McCammon:2002,Hickox:2006,Lehmer:2012} (CXB).  \citet{Fialkov:2017} noted that when one subtracts the sources at $z \lesssim 10$ considered by \citet{Cappelluti:2012}, the X-ray flux from the Cosmic Dawn should not exceed  $J_{0.5-2\text{keV}} \approx 2.5 \times 10^{-13}$\,erg\,s$^{-1}$\,cm$^{-2}$\,keV$^{-1}$\,deg$^{-2}$ over the $0.5-2$\,keV band. We compute $\epsilon_\nu$ 
from our black hole population through 
\begin{equation}
\epsilon(\nu,z) = \int d\text{M}_{\text{bh}} n(\text{M}_{\text{bh}}) L_\nu(\text{M}_{\text{bh}})
\end{equation}
where $n(\text{M}_{\text{bh}})$ is the number of black holes of a given mass, and $L_\nu(\text{M}_{\text{bh}})$ is the luminosity of emission from the black hole at frequency $\nu$ 
. We extrapolate the $0.1-2$\,keV X-ray luminosities from \citet{Wang:2006} to harder energies (which are redshifted into the 0.5-2 keV band by $z=0$) by assuming an X-ray spectral index of $0.9$, typical of AGN \citep{Nandra:1994,Reeves:2000,Piconcelli:2005,Page:2005}, and a minimal X-ray energy of $0.5$\,keV due to self-absorption by the interstellar medium in each black hole's host galaxy \citep{Mesinger:2013,Fialkov:2014,Pacucci:2014,Das:2017}. We report $J_{0.5-2\text{keV}}$ versus $f_{\text{edd}}$ in Fig.~\ref{fig:Xrays}.  We find that our models mostly lie below the X-ray background constraint. 

\subsection{Black Hole Densities}\label{ssec:Density}
\begin{figure*}
\includegraphics[width=\textwidth]{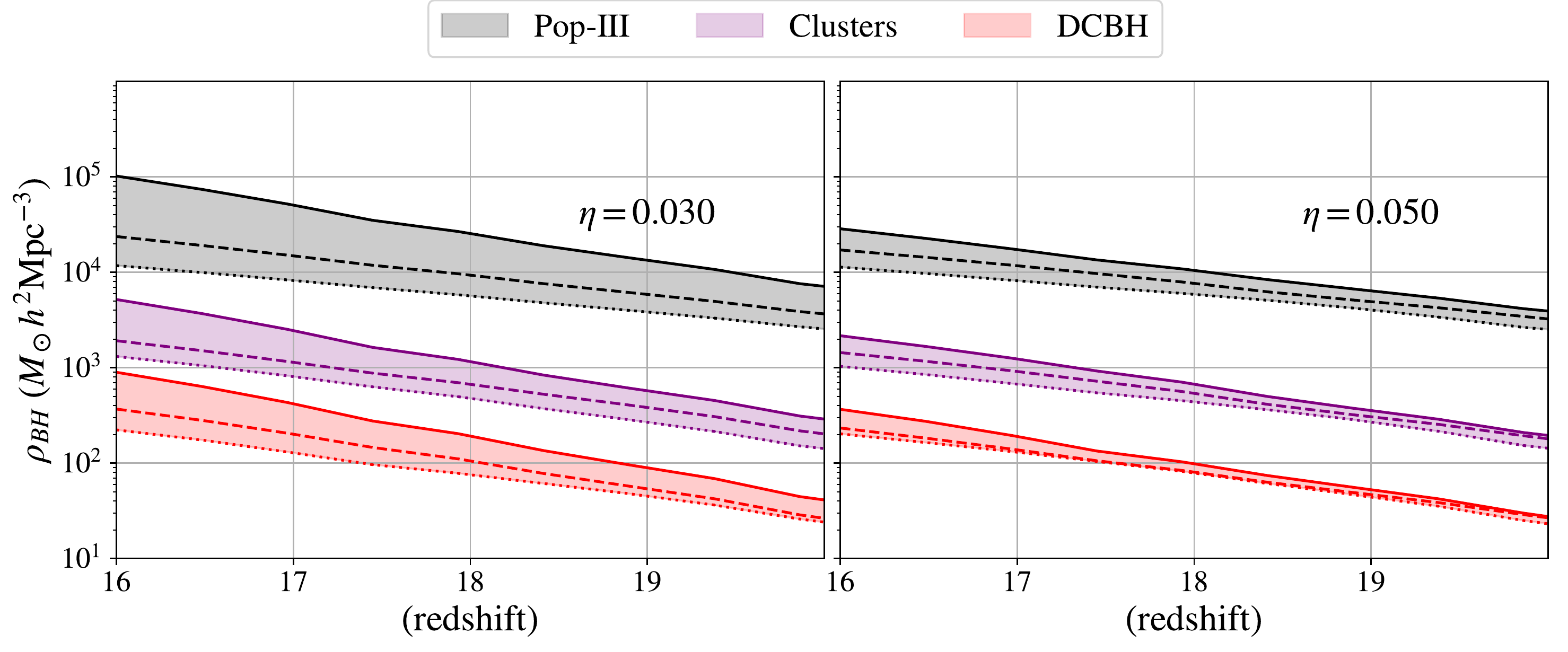}
\caption{ The co-moving density of black holes produced by our formation model. In all cases, they are below the limit determined from dynamical black hole masses in the local universe ($\approx 1.1 \times 10^{6} \text{M}_\odot h^2\text{Mpc}^{-3}$), though allowing for accretion to proceed unregulated would violate these constraints at lower redshifts. Shaded regions indicate densities obtaind between the Eddington ratios of $10^{-2}$ and $1$ for each set of models.}\label{fig:Density}
\end{figure*}

We next inspect the density of black holes produced by our accretion/emission models in Fig.~\ref{fig:Density}. By redshift 16, the most optimistic Pop-III models, which might produce or over-produce the EDGES signature are approaching the limit of $\approx 1.1 \times 10^{6} \text{M}_\odot h^2\text{Mpc}^{-3}$ determined from dynamical black hole masses in the local Universe \citep{Merritt:2001}. It follows that if these accreting sources are responsible for the EDGES amplitude, then their growth and emission must be curtailed through some feedback process at lower redshifts. 

\subsection{Contribution to Reionization}\label{ssec:Ionization}
Seeing that it is possible for our Pop-III model to produce a sufficient radio background to explain the amplitude of the EDGES absorption feature, we investigate this model's implications on reionization. From each quasar's X-ray luminosity at $2$\,keV, we assign an optical luminosity at $2500\,\text{\AA}$ using the relationship observed by \citet{Lusso:2010},
\begin{equation}\label{eq:UV}
\log_{10} L_{2500\text{\AA}} = \log_{10} L_{2\text{keV}} + 2.605 \times \alpha_{ox},
\end{equation}
where we use their empirically derived value of $\alpha_{ox}=1.37$. We compute the ionizing flux by extrapolating from $2500\,\text{\AA}$ to $912\,\text{\AA}$ using a power law of $-0.65$ and assume a steeper spectrum blue-ward of $912\,\text{\AA}$ with a spectral index of $-1.7$ \citep{Lusso:2015}. We compute the rate of Hydrogen-ionizing photons as $\dot{n}_{\text{ion}}(z)=f_{\text{esc}}\int \frac{d \nu}{h_P\nu}\epsilon[\nu,z]$, 
where $h_P$ is Planck's constant and $f_{\text{esc}}$ is the fraction of ionizing photons that are able to escape into the IGM.

The escape fraction of high redshift galaxies remains highly uncertain. \citet{Ma:2015} find escape fractions between 0.01 and 0.05 for $\sim 10^{9} \text{M}_\odot$ halos in simulations, while observations of larger galaxies at lower redshifts find escape fractions $f_{\text{esc}} \sim 10^{-2}-0.5$ \citep{Bridge:2010,Izotov:2016,Vanzella:2016,Vasei:2016,Grazien:2017}. These may not be representative for our Pop-III hosts which have masses of $\approx 10^6 \text{M}_\odot$, when our seeds are formed, and grow to between $\approx 10^8 \text{M}_\odot$ and $\approx 10^9 \text{M}_\odot$ by redshift 16. Theoretical models of quasar-driven reionization often assume that the escape fraction of galaxies hosting quasars was significantly higher than their non-quasar hosting counterparts and take $f_{\text{esc}} \sim 1$ (e.g. \citet{Madau:1996,Madau:2015}). Despite this, recent observations by \citet{Micheva:2017} find $f_{\text{esc}}\lesssim10^{-2}$ for AGN at $z \approx 3$. 


In order to prevent X-rays from escaping and heating the IGM, any black holes producing the EDGES feature would need to have $f_\text{esc} \sim 0$ (see the discussion in \S~\ref{sec:Discussion}). However, it is also possible that the black holes were not as obscured as our heating constraint suggests, but were extraordinarily radio loud instead, overcoming any simultaneous heating. Thus, to be conservative, we consider  escape fractions between $0.05$ and $0.2$, much larger than the value imposed by our heating constraint but more in line with observations of lower-redshift sources.

Following \citet{Madau:2015}, we assume that photons between $1$ and $4$\,Ryd contribute primarily to Hydrogen ionization while photons above $4$\,Ryd contribute to Helium ionization. We then obtain the volumetric filling fractions of H$_{\text{II}}$ ($\text{Q}_{\text{HII}}$), and He$_{\text{III}}$ ($\text{Q}_{\text{HeIII}}$) by integrating the ionization equations (see \citet{Madau:2015} and references therein),
\begin{align}
\dot{\text{Q}}_{\text{HII}} &= \frac{\dot{n}_{\text{ion}}}{\langle n_{\text{H}}\rangle} - \frac{\text{Q}_{\text{HII}}}{t_{\text{rec,H}}}\\
\dot{\text{Q}}_{\text{HeIII}} &= \frac{\dot{n}_{\text{ion}}}{\langle n_{\text{He}}\rangle} - \frac{\text{Q}_{\text{HeIII}}}{t_{\text{rec,He}}},
\end{align}
where $\langle n_{\text{H}} \rangle$ is the co-moving number density of Hydrogen atoms and $\langle n_{\text{He}} \rangle$ is the co-moving number density of Helium. We use these authors' expressions for the recombination-times of Hydrogen, $t_{\text{rec,H}}$, and Helium, $t_{\text{rec,He}}$, and halt the evolution of $\text{Q}_{\text{HII/HeIII}}$ at $z$ such that $\text{Q}_{\text{HII/HeIII}}=1$ or $z=16$, when we stop growing our black holes; whichever occurs first. 

\begin{figure*}
\includegraphics[width=\textwidth]{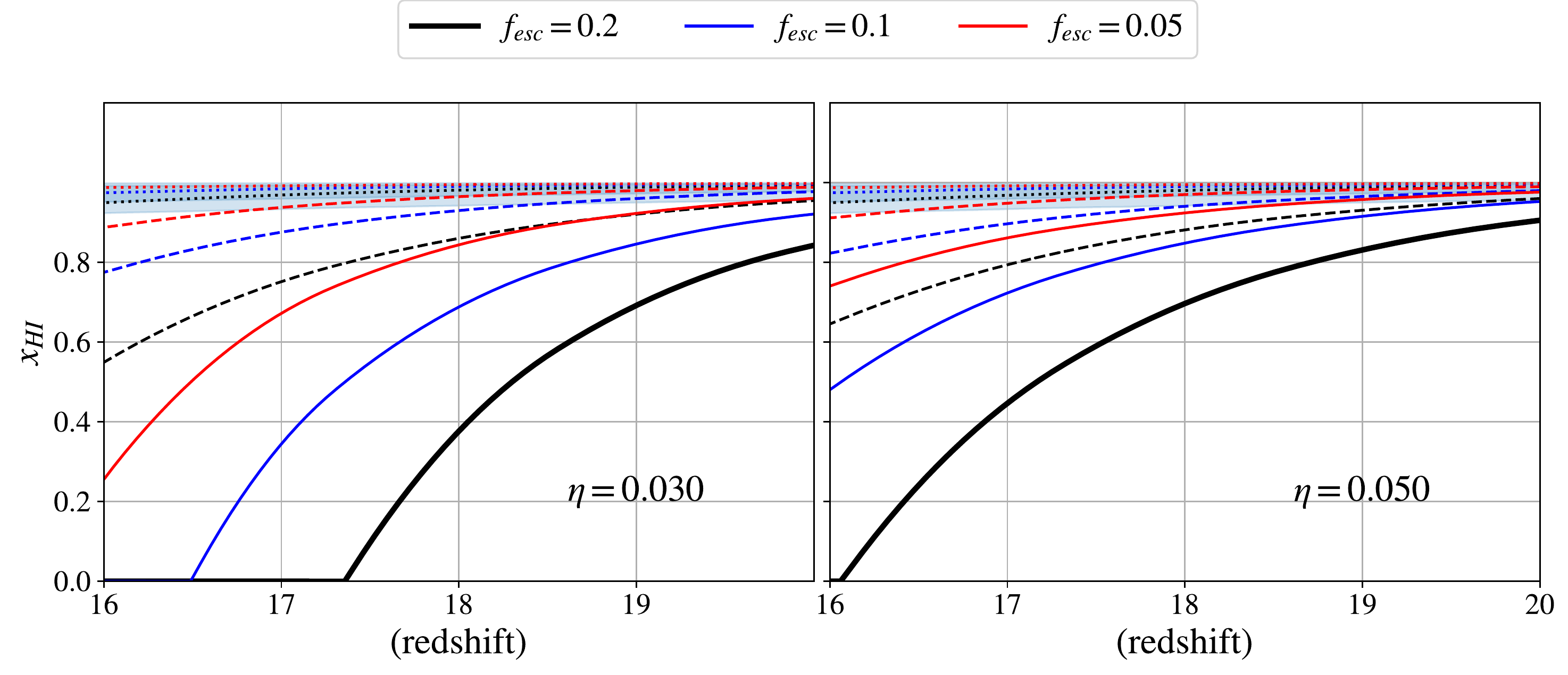}
\caption{The evolution of the Hydrogen neutral fraction, $x_{\text{HI}}$, for the ionizing flux of our Pop-III model and escape fractions of $0.2$ (black lines), $0.1$ (blue lines), and $0.05$ (red lines). Dotted lines correspond to $f_\text{edd} = 10^{-2}$, dashed lines correspond to $f_{\text{edd}}=10^{-1}$, and solid lines correspond to $f_{\text{edd}}=1$.
The models that are able to produce the EDGES absorption feature also reduce the neutral fractions at $z=17$. In particular, for $\eta=0.03$ (left panel) and $f_{\text{edd}}=1$, the neutral fraction is $\approx0.7$ for $f_{\text{esc}}=0.05$ and $\lesssim0.4$ for $f_{\text{esc}}=0.1$. The blue shaded regions denote the 1 and $2\,\sigma$ constraints on $x_{\text{HI}}$ from \citet{Millea:2018}. Demanding that the IGM remains unheated implies $f_\text{esc} \approx 0$, far lower than the models considered in this plot. }\label{fig:xHI}
\end{figure*}

In Fig.~\ref{fig:xHI}, we show the evolution of the neutral fraction, $x_{HI}=1-\text{Q}_{\text{HII}}$, with redshift for three different values of the escape-fraction, $f_{\text{esc}}=0.2$, $f_{\text{esc}}=0.1$, and $f_{\text{esc}}=0.05$. 

For $f_{\text{esc}}=0.1$, all of our models introduce non-negligible amounts of ionization by $z=17$, with $x_{HI}$ ranging from $0.4$, for our most emissive model, to $\approx 0.9$ at lower Eddington ratios. While this means that the overall amplitude of the absorption feature would fall by a factor of $\approx 2-4$, these very emissive models produce enough excess radio emission (Fig.~\ref{fig:Boost}) to make up for such a reduction. We also see that increasing $f_{\text{esc}}$ to $\approx 0.2$ can cause ionization to occur too early, erasing the absorption feature. 

We conclude that with $f_{\text{esc}} \lesssim 0.1$, the ionizing flux is not problematic for the amplitude of the absorption feature. Furthermore, it is possible that fine-scale, optically thick substructures, such as Damped Lyman-$\alpha$ systems, can delay ionization further by $\Delta z\approx 2$ \citep{Sobacchi:2014}. 

\begin{figure*}
\includegraphics[width=\textwidth]{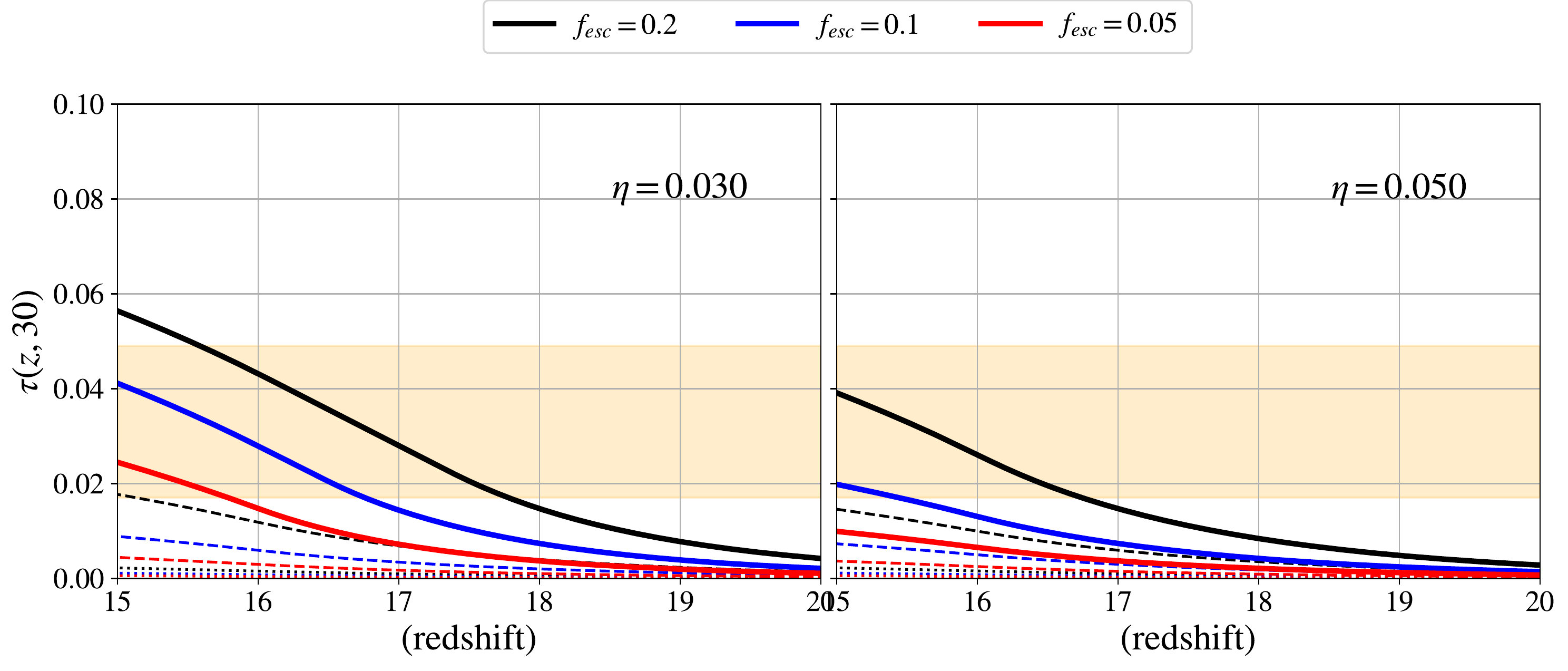}
\caption{
The integrated optical depth between $z$ and $z=30$ predicted by our Pop-III black hole model assuming that accretion shuts down at $z=16$ and escape fractions $f_{\text{esc}}=0.05$ (red lines), $f_{\text{esc}}=0.1$ (blue lines) and $f_{\text{esc}}=0.2$ (black lines). Dotted lines 
correspond to $f_{\text{edd}}=10^{-2}$ 
, dashed lines correspond to $f_\text{edd} = 10^{-1}$, and solid lines correspond to $f_{\text{edd}}=1$. The orange shaded region denotes the range of $\tau(15,30)$ allowed by \citet{Heinrich:2018}. The obscuration necessary for keepting the IGM unheated for the duration of the trough ($\sim 100$\,Myr) would naturally lead to $f_\text{esc}$ values much smaller than any considered in this plot. }\label{fig:Tau}
\end{figure*}

While moderate ionization at high redshifts can still allow for the EDGES amplitude, it might come into conflict with the Thomson scattering optical depth of CMB photons measured by \citet{Planck:2016}.
Rather than calculate $\tau$ to $z=0$ from our black holes, which would only contribute with part of the total ionization that is mainly expected from star formation at lower redshifts, we calculate the integrated optical depth between $z$ and $z_{\text{max}}=30$, $\tau(z,30)$, for which Planck polarization constraints of $\tau(15,30)=0.033\pm0.016$ have recently been derived \citep{Heinrich:2018}. We show the results in Figure~\ref{fig:Tau} 
. Our $f_\text{esc} \approx 0.1$ models tend to be consistent with \citet{Heinrich:2018}'s results though more recent work by \citet{Millea:2018} (whose $x_\text{HI}$ predictions we show in Fig.~\ref{fig:xHI}) arrive at more stringent constraints, which would only allow for $f_\text{esc} \lesssim 0.05$ if accretion was Eddington limited.  

\begin{figure*}
\includegraphics[width=\textwidth]{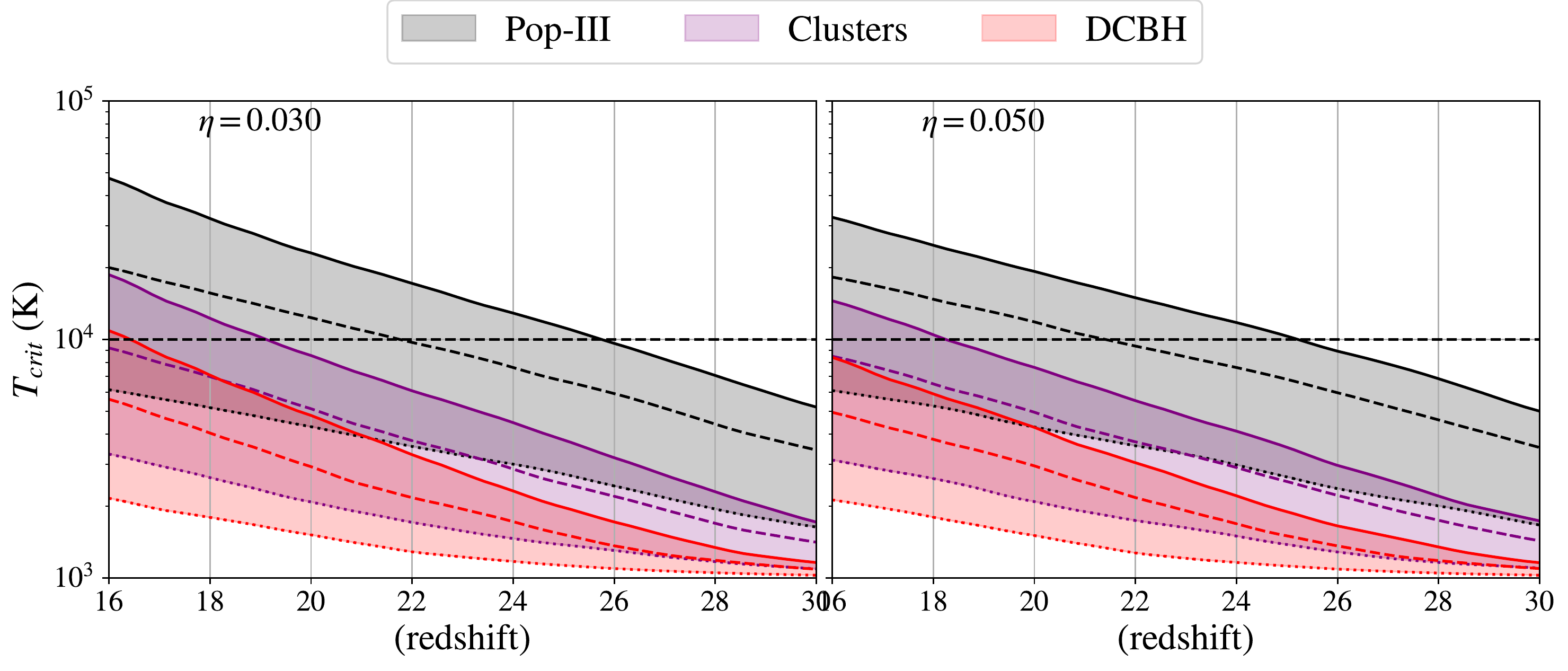}
\caption{The critical virial temperature of dark-matter halos below which the background of Lyman-Werner photons emitted by black holes prevents H$_2$ cooling. Dotted, dashed, and solid lines denote $f_\text{edd} = 10^{-2},10^{-1}$, and $10^0$ respectively. A wide range of parameters in our Pop-III black hole model produces enough of a Lyman-Werner background to shut-down star formation through $H_2$ cooling in all molecular cooling halos at $z \gtrsim 20$ which might serve as a natural regulation mechanism that prevents the over-production of black holes. 
}\label{fig:LW}
\end{figure*}


\subsection{Lyman-Werner Feedback}\label{ssec:LWBackground}

Unlike their ionizing counterparts, Lyman-Werner (LW) photons between 11.2 and 13.6\,eV are capable of escaping into the IGM and dissociating the H$_2$ molecules necessary for the cooling and collapse of baryons onto mini-halos \citep{Haiman:1997}. Thus, a sufficiently strong LW background generated by early black-holes can provide a natural feedback mechanism for scenarios where the black-hole seeds form from Population-III stars.

We use the UV SED described in \S~\ref{ssec:Ionization} to predict the Lyman-Werner background generated by the accreting black holes as a function of redshift using equation~\ref{eq:Intensity}, with the upper integration limit of $z_m$, where $ z_m + 1 = 1.04 \times (z+1)$ to account for the fact that LW photons can only redshift by a factor of $\approx 4\%$ before being absorbed by a Lyman transition \citep{Visbal:2014}. In Fig.~\ref{fig:LW}, we plot the Virial temperature corresponding to the minimum halo mass that can provide enough self-shielding to allow for H$_2$ cooling to occur within a Hubble time, $T_\text{crit}$, approximated by Equation~4 in \citet{Visbal:2014}. By redshift $\approx 24-26$, many of the models capable of generating the EDGES feature bring the minimum halo mass above the atomic cooling threshold, which would severely curtail Pop-III seed formation, serving as a potential  mechanism for shutting down black-hole production before $z \approx 16$ and satisfying
constraints on black-hole production and radiative backgrounds. Pop-III shutdown from Black-hole generated LW feedback before $z \approx 20$ 
stands in contrast to models in which the LW background only comes from Pop-III stars themselves. Scenarios without black-holes have been found to allow for star formation in pristine molecular cooling halos until $z \lesssim 10$ \citep{Jaacks:2018,Mebane:2018}. The formula from \citet{Visbal:2014}, that we have used to predict $T_\text{crit}$, ignores the potentially countervailing feedback from X-rays. Since X-rays generate free electrons, they work to catalyze H$_2$ formation and decrease $T_\text{crit}$. The actual shut-down redshift for Pop-III formation is therefore somewhat lower than our simple analysis predicts and we defer a more detailed analysis to future work.

\subsection{Contribution to the Cosmic Infrared Background}
We also check our model's contribution to the Cosmic Infrared background at 3.6\,$\mu$m using a procedure similar to what is described in \S~\ref{ssec:Ionization}. We assume that our black holes have a luminosity at $2500$\,\AA\, given by equation~\ref{eq:UV} with a spectral index of $-0.65$
redward of $912\,\AA$. For this treatment, all of our models predict IR fluxes at $3.6\,\mu$m (integrated across the IRAC filter width of $\approx 1\,\mu$m) that are $\lesssim 10^{-4}$\,nW\,m$^{-2}$\,Hz$^{-1}$\,sr$^{-1}$. This is well below the typical values that are supposedly from Cosmic Dawn sources, of $\sim 1$\,nW\,m$^{-2}$\,sr$^{-1}$ \citep{Kashlinsky:2007}.  
This calculation ignores re-processed IR emission caused by the absorption of UV and X-ray photons by any obscuring clouds.  We leave the study of re-processed radiation, which might greatly exceed the unobscured quasar emission considered here, for future investigations.

\section{Discussion}\label{sec:Discussion}
While we have shown that black holes can produce a CMB-level 21\,cm background by $z\approx 17$, under optimistic assumptions of accretion rates and present day correlations between X-ray and radio luminosities, there are still significant issues that a black hole driven model must overcome before it is considered as a serious explanation for the EDGES excess.

Firstly, our accretion model produces copious amounts of X-rays which would raise the temperature of the IGM (and potentially contribute significantly to re-ionization). Any such rise in $T_s$ would erase the gains made in the amplitude of absorption unless energy deposition of the X-rays were delayed until redshift $z \lesssim 16$, where EDGES observes the rise out of the trough. \citet{Fialkov:2014} (also see \citet{Mesinger:2013}) discuss a heating scenario in which a hard X-ray spectrum due to obscuration at $\lesssim 1$\,keV could reduce their energy and delay energy deposition, significantly enhancing the amplitude of the absorption feature. Thus, one possibility is that the progenitors to super-massive black holes were born in heavily obscured environments that prevented the escape of X-ray and UV photons but still allowed for Ly-$\alpha$ to strongly couple the gas to the adiabatically cooled IGM. \citet{Tanaka:2016} come to a similar conclusion as they notice that unobscured X-ray emission from Pop-III black holes would erase the absorption feature.

We may estimate the required obscuration to keep the IGM unheated over the duration of the EDGES feature by considering the (physical) mean-free path of an X-ray in the IGM:

\begin{equation}
\lambda_X \approx 1.56 x_\text{HI}^{-1}\left(\frac{1+z}{17}\right)^{-3} \left(\frac{E}{500\text{eV}}\right)^{2.6} \text{Mpc}.
\end{equation}

If X-rays with energies below $E_\text{min}$ are absorbed locally, energy deposition (and heating) of the IGM will be delayed by $\tau_X = \lambda_X(E_\text{min})/c$. 

 Provided that the gas remains mostly unheated over the duration of the absorption trough, and the up-turn at $z \approx 14$ is driven by rapid heating, the shape and duration of the absorption trough reported by EDGES actually provides us with constraining information on the local column depths of black holes providing the radio background. The width of the trough is $\approx 100$\,Myr. Solving for $ c \times 100\,\text{Myr} = \lambda_X(E_\text{min})$ yields $E_\text{min} \approx 1.7\,\text{keV}$. The column depth required to obscure X-rays up to $E_\text{min} \approx 1.7 \,\text{keV}$ can be obtained by setting $N_\text{H} \sim \sigma_\text{H}^{-1}(1.7 \text{keV}) \sim 5 \times 10^{23} \text{cm}^{-2}$, where $\sigma_\text{H}$ is the photo-ionization cross section of Hydrogen. Models explaining the X-ray background from AGN at lower redshift often invoke similar obscuration column depths \citep{Fabian:1999a}, requiring that the gas forms a dense, efficiently cooled spheroid around the black hole. At high redshift, the obscuration eventually might be disrupted by the black hole's emission, once it has grown to a sufficiently large size, which might hasten the evolution of $T_s$ and explain the very rapid elimination of the absorption feature by $z=14$. We check that such column depths would not obscure radio emission by assuming that the gas with a fixed column depth forms a spherical cloud with constant density around the black hole. Even if the gas were highly ionized, we find that the plasma frequencies for such column depths do not exceed $\approx 10$\,MHz.

Our requirement that the IGM remains unheated for $\approx 100$\,Myr after black-holes start emitting imposes the condition that

\begin{equation}
f_\text{esc}(N_\text{H} \sim 10^{23})  \sim \exp \left[ 10^{23} \text{cm}^{-2} \sigma_\text{H}(13.6 \text{eV}) \right] \approx 0.
\end{equation}

Thus, imposing obscuration requirements based on the timing of the signal guarantees that the AGN do not contribute to reionization until the obscuring gas is cleared away, allowing them to obey the Planck constraint on $\tau$.

Since our black-hole seed masses are in the range that would form from the direct collapse of a star (without a supernovae), radiative feedback might be low enough to allow for high obscuration (and efficient growth) of the black hole. At the same time, without any heating, our most optimistic accretion scenario over-predicts the amplitude of the EDGES feature by a factor $\approx 5$, so there are scenarios in which some heating might be tolerated while still recovering the anomalously large absorption feature reported by \citet{Bowman:2018}. 

Secondly, the models that are capable of producing the reported EDGES amplitude assemble a large fraction of the co-moving black hole mass at $z=0$ (between $1$ and  $10$ percent) by $z\approx 17$, and would likely over-produce the black hole density if exponential growth were allowed to proceed unregulated. Feedback mechanisms will have to be invoked to suppress the formation of future Pop-III seeds and regulate the growth of the existing black holes beyond $z \approx 16$. Fortunately, it is expected that Pop-III formation naturally ends once the metallicity of gas reaches $\approx 10^{-6}-10^{-3}Z_\odot$ \citep{Bromm:2001,Omukai:2005} and while a number of models show that Pop-III formation can continue down to redshifts of $z \lesssim 10$, we have shown that the additional Lyman Werner feedback generated by rapidly accreting Pop-III black holes capable of producing the EDGES feature is large enough to  shutdown black hole seed formation at $z \sim 20$.

Obscuration of the Pop-III black holes necessary to prevent significant heating and ionization in the IGM would naturally lead to large amounts of energy being deposited within their local environments, providing an additional regulation mechanism. Detailed modeling to determine whether a combination of obscuration and feedback can both delay heating and shut down accretion in a manner that explains the EDGES absorption feature is the subject of future work.

Thirdly, while only our Pop-III model appears to produce a sufficiently bright radio background to explain the EDGES feature, detailed simulations have found that heating and photo-ionization feedback tend to prevent Pop-III remnants from achieving high accretion rates (e.g. \citet{Alvarez:2009}). Hence, rapid accretion within atomic cooling halos 
might actually be more plausible 
than in our Pop-III scenario (for example, models in which $\sim 100$\,$\text{M}_\odot$  seeds experience super-critical accretion from a thick, obscuring, and radiative inefficient torus \citep{Volonteri:2005}). An argument for atomic cooling halos can also be raised from the timing of the absorption feature, which may start at too low a redshift to be caused by molecular cooling halos \citep{Kaurov:2018}. Thus scenarios where black-hole seeds formed in atomic cooling halos in higher abundances then what is predicted in the models that we based our DCBH and cluster scenarios on are an interesting possibility that should be explored further. 

While a radio background might explain the EDGES feature, mechanisms different from those explored here have been suggested for producing this background. They include cosmic rays accelerated in supernovae \citep{Mirocha:2018} and instabilities in mini-clusters of Axionic dark matter \citep{Fraser:2018}. A black-hole driven scenario might be distinguished from a scenario involving star-forming Galaxies in several ways. 

Firstly, Eddington accretion can lead to exponential time-evolution of radiative emissivities, significantly faster than scenarios driven by star-forming galaxies, where emissivities evolve proportional to the star-formation rate. Secondly, X-ray emission driven from inverse bremsstrahlung of the same cosmic rays producing the radio emission is known to be significantly softer than X-ray emission from black-hole accretion, which would lead to rapid heating and higher-contrast spatial fluctuations \citep{Pacucci:2014}.

Spatial HI intensity mapping and deep point-source surveys might also be used to distinguish black-hole radio emission from a background generated by collapsing axionic mini-clusters. In the axion model, the radio background could be generated by halos significantly below the atomic cooling threshold. Since collapsing axionic halos would not generate X-ray emission, the timing and spatial variation between hot/cold patches of HI and the radio background would be decoupled and might help to confirm or reject such a model. In addition, radio emission from axionic decays might be distinguished by the fact that such emission is intrinsically narrow band. 

\begin{figure}
\includegraphics[width=.5\textwidth]{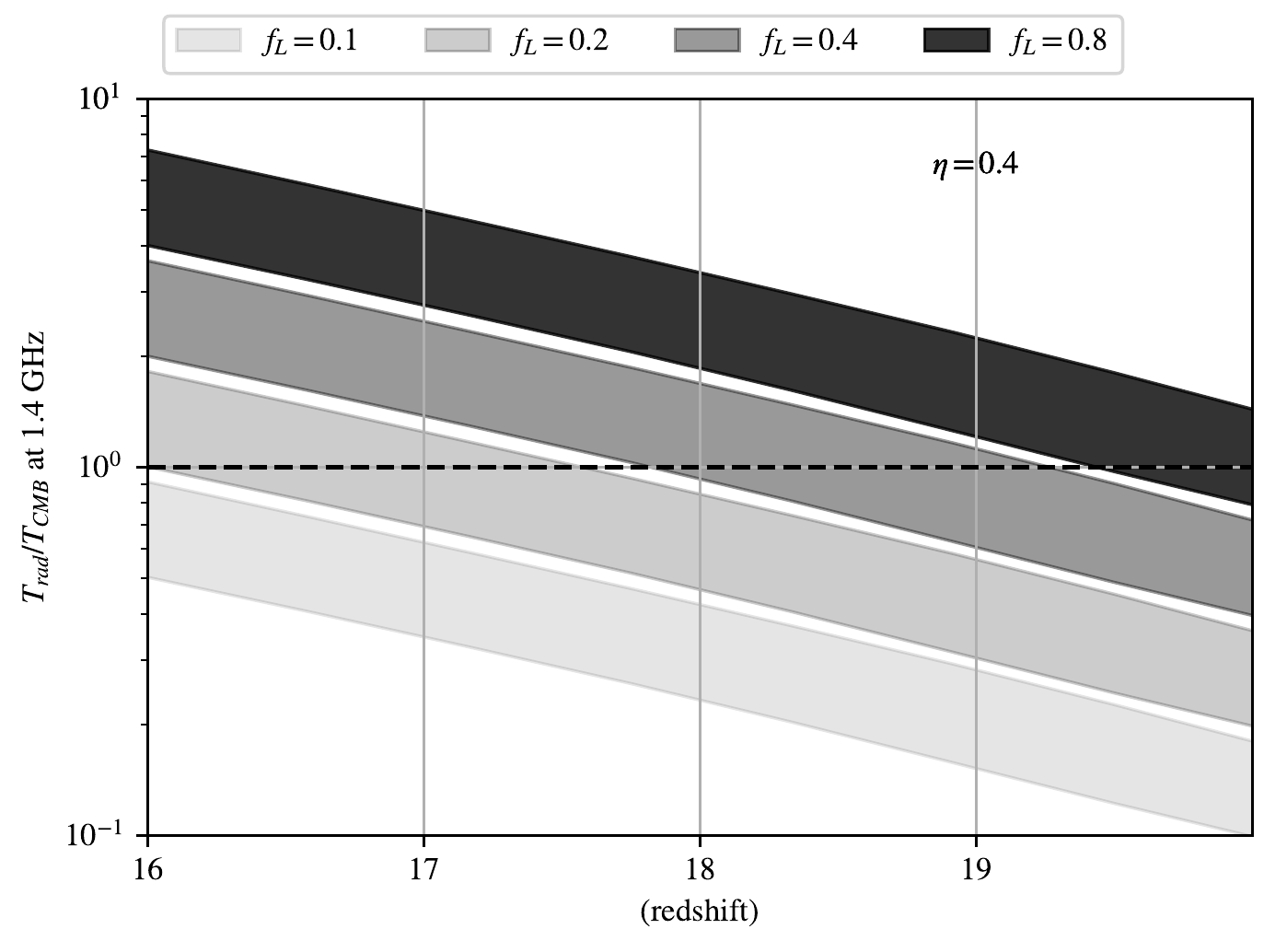}
\caption{The same as Fig.~\ref{fig:Boost} but now only considering Pop-III black holes and fixing $\eta=0.4$ 
, which is an upper limit for thin-disk accretion scenarios with black holes spinning close to their theoretical maximum. Filled regions correspond to $\text{T}_\text{rad}/\text{T}_\text{CMB}$ between $f_\text{edd} = 0.1$ and $f_\text{edd} = 1$ with fixed $f_L$. Doubling the radio-loud fraction from what is observed at the present day (i.e., using $f_L=0.2$) gives us enough emission to produce the EDGES absorption feature.}\label{fig:VaryFL}
\end{figure}

In our study, we restricted ourselves to scenarios with very rapid accretion driven by low radiative efficiencies ($\eta < 0.1$).
Within the thin disk paradigm \citep{Shakura:1973}, this corresponds to a black-hole spin of $a/\text{M}_\text{bh} \lesssim 0.7$. It is currently unknown what the distribution of initial spins for super-massive black hole progenitors might be and in what fashion they were modified by accretion. Spin evolution through accretion depends heavily on whether accreting material follows pro-grade or retro-grade orbits and whether accretion episodes are coherent \citep{King:2008}. Higher radiative efficiencies than what we examined might still explain the EDGES feature but would require larger radio luminosities per black hole or larger radio-loud fractions than we examined. In Fig.~\ref{fig:VaryFL} we show $T_\text{rad}/T_\text{CMB}$ for $\eta = 0.4$ 
, close to its theoretical maximum, for different values of $f_L$ and $f_\text{edd}$. In this radiatively efficient scenario, $f_L=0.1$ cannot produce a background large enough for the EDGES feature, but simply doubling the radio-loud fraction (or equivalently the expectation value of radio luminosity) to $f_L=0.2$ can.

\section{Conclusion}\label{sec:Conclusions}
In this paper, we have investigated the plausibility that the absorption feature recently detected by the EDGES experiment might have been produced by an additional radio background arising from accretion onto growing black holes during the Cosmic Dawn. To do this, we combined low-redshift empirical relationships between AGN X-ray luminosities and radio emission, with a semi-analytic framework that creates new seed black holes based on the halo mass function and grows them exponentially through Eddington limited accretion. With this framework, we have explored plausible radio backgrounds over a broad range of physically motivated seed populations and growth rates. Our main conclusions from this study are:

\begin{enumerate}
\item Black holes forming and growing at physically plausible rates 
can produce a radio background sufficient to explain the amplitude of the EDGES absorption feature while satisfying existing constraints on the soft X-ray background and faint radio source counts. 

\item By demanding that the IGM remain unheated over the duration of the EDGES trough ($\sim 100$\,Myr), we find these black holes would need to be obscured with column depths of $N_\text{H} \approx 5 \times 10^{23}$\,cm$^{-2}$. Such large column depths would also prevent these black holes from ionizing the IGM too early.

\item Of the models we considered, none that are limited by the atomic cooling threshold reached more than $10\%$ of the radio emission needed to explain the EDGES feature. Increasing the DCBH fraction by $\sim 100$ (through enhancements such as baryon-dark matter velocity offsets) could explain the EDGES feature but would come into conflict with the faint source counts constraint in Fig.~\ref{fig:Radio}, though a steeper spectral index or reducing $f_\text{seed}$ in exchange for a further increase in $f_\text{halo}$ can mitigate this issue.  The cluster collapse scenario we considered is still viable if a $\sim 10 \times$ larger fraction ($\approx 50\%$) of halos hosted such seeds. It is also possible that a black-hole population arising from a combination of all three mechanisms in a wide range of halo masses could produce a sufficiently bright radio background. 

\item In order to avoid over-producing the local Black-hole population and radiative backgrounds at low redshift, the emission and growth of these black holes would need to be curtailed at $z \lesssim 16$. We have shown that the Lyman-Werner background generated by the black-holes themselves is sufficient to shut down star formation in molecular cooling halos at $z lesssim 20$, potentially providing the necessary feedback mechanism. 
\end{enumerate}
While we have diligently checked the implications of our model against observational constraints of radio counts, the CXB, CIB, and CMB optical depth, we caution our readers to heed the caveats in \S~\ref{sec:Discussion}.


\acknowledgments
\section{Acknowledgements}
The authors thank the anonymous referee, Judd Bowman, Adam Lidz, Rennan Barkana, Anastasia Fialkov, and Jordan Mirocha for useful and insightful comments along with Florian Bolgar for correcting several numerical errors. Calculations in this work were performed using the {\tt Colossus} library \citep{Diemer:2017}. AEW’s contribution was supported by an appointment to the NASA Postdoctoral Program at the California Institute of Technology Jet Propulsion Laboratory. Part of the research was carried out at the Jet Propulsion Laboratory, California Institute of Technology, under a contract with the National Aeronautics and Space Administration. R.A.M. was supported by the NASA Solar System Exploration Virtual Institute cooperative agreement 80ARC017M0006.

\bibliographystyle{apj}
\bibliography{BH_Paper}


\end{document}